\documentclass[aip,jmp,amsmath,amssymb,reprint]{revtex4-1}

\usepackage{graphicx}% Include figure files
\usepackage{dcolumn}% Align table columns on decimal point
\usepackage{bm}% bold math
\usepackage[utf8]{inputenc}
\usepackage[T1]{fontenc}
\usepackage{etoolbox}
\usepackage{xcolor}
\usepackage{tabularray}
\usepackage{graphicx}% Include figure files
\usepackage{dcolumn}% Align table columns on decimal point
\usepackage{bm}% bold math

\usepackage[a4paper, total={6in, 9in}]{geometry}\usepackage{graphicx}
\usepackage{amssymb}
\usepackage{amsthm}
\usepackage{bbm}
\usepackage{bm}
\usepackage{amsmath}
\usepackage{mathrsfs}
\usepackage{amssymb}
\usepackage{exscale}
\usepackage{easybmat}
\usepackage{xcolor}
\usepackage{comment}
\usepackage[normalem]{ulem}
\usepackage{booktabs}
\usepackage{multirow}
\usepackage{siunitx}
\usepackage[figurename=Figure, labelsep=period, font=footnotesize, labelfont=bf, format=plain]{caption}
\usepackage{titlesec}

\newcommand{\ii}{\mathrm{i}}

\usepackage{xcolor}
\definecolor{custom}{RGB}{0,144,226}
\newcommand{\Revision}[1]{{\color{black} #1}} 

\makeatletter
\def\@email#1#2{%
 \endgroup
 \patchcmd{\titleblock@produce}
  {\frontmatter@RRAPformat}
  {\frontmatter@RRAPformat{\produce@RRAP{*#1\href{mailto:#2}{#2}}}\frontmatter@RRAPformat}
  {}{}
}%
\makeatother
\begin{document}

%\preprint{AIP/123-QED}

\title[]{Mechanical intelligence via fully reconfigurable elastic neuromorphic metasurfaces}
% Force line breaks with \\
\author{M. Moghaddaszadeh}
\altaffiliation[These authors contributed equally to this work]{}
\affiliation{Department of Mechanical and Aerospace Engineering, University at Buffalo (SUNY), \\Buffalo, NY 14260-4400, USA}
\affiliation{Department of Civil, Structural and Environmental Engineering, University at Buffalo (SUNY), Buffalo, NY 14260-4300, USA}

\author{M. Mousa}%
\altaffiliation[These authors contributed equally to this work]{}
\affiliation{Department of Mechanical and Aerospace Engineering, University at Buffalo (SUNY), \\Buffalo, NY 14260-4400, USA}

\author{A. Aref}
\affiliation{Department of Civil, Structural and Environmental Engineering, University at Buffalo (SUNY), Buffalo, NY 14260-4300, USA}

\author{M. Nouh}
\altaffiliation[Corresponding author: \url{mnouh@buffalo.edu}]{}
\affiliation{Department of Mechanical and Aerospace Engineering, University at Buffalo (SUNY), \\Buffalo, NY 14260-4400, USA}
\affiliation{Department of Civil, Structural and Environmental Engineering, University at Buffalo (SUNY), Buffalo, NY 14260-4300, USA}

\date{\today}% It is always \today, today,
             %  but any date may be explicitly specified

\begin{abstract}
The ability of mechanical systems to perform basic computations has gained traction over recent years, providing an unconventional alternative to digital computing in off grid, low power, and severe environments which render the majority of electronic components inoperable. However, much of the work in mechanical computing has focused on logic operations via quasi-static prescribed displacements in origami, bistable, and soft deformable matter. In here, we present a first attempt to describe the fundamental framework of an elastic neuromorphic metasurface that performs distinct classification tasks, providing a new set of challenges given the complex nature of elastic waves with respect to scattering and manipulation. Multiple layers of reconfigurable waveguides are phase-trained via constant weights and trainable activation functions in a manner that enables the resultant wave scattering at the readout location to focus on the correct class within the detection plane. We further demonstrate the neuromorphic system's reconfigurability in performing two distinct tasks, eliminating the need for costly remanufacturing. 
\end{abstract}

\maketitle
%%%%%%%%%%%%%%%%%%%%%%%%%%%%%%%%%%%%%%%%%%%%%%%%%%%%%%%%%%%%%%%%%%%%%%%%%%%%%%%%%%%%%%%%%%%%%%
\section{\label{sec:intro}INTRODUCTION}
Mechanical computing \cite{Hiromi2010}, a research field older than electronic computing, has gained a lot of interest over the past few years. Despite the limited capabilities of mechanical computers, the ability to integrate accurate computational and morphological capabilities with minimal energy requirements in a self-contained mechanical structure remains invaluable \cite{song2019additively}. The recent surge in mechanical computing research has been motivated by new and transformative technologies requiring rapid (yet simple) data processing, and directly encoded autonomy and intelligence \cite{riley2022neuromorphic}. The latter presents an opportunity to exploit the dynamic behavior of structures, and capitalize on certain synergies associated with physical phenomena such as material response, deformation, and scattering to accomplish effective computation with minimal resources \cite{DARPA_NAC}. In keeping with these needs, several mechanical concepts have recently emerged with the goal of performing combinational logic \cite{Helou2022} and basic mathematics \cite{zuo2018, zuo2022}, using for example conductive polymers \cite{Helou2021} or bistable spring-mass chains \cite{Ion2017, Bilal2017} that propagate mechanical signals when triggered. In tandem, there have been reports of origami-inspired metamaterials in which shear and expansion responses along different directions change the configuration state \cite{Treml2018, MENG2021, Liu2023}, and snap-through mechanisms where elastic instabilities are exploited to enable the creation of logic gates \cite{Mei2021}.

In pursuit of systems capable of learning and adaptation, structures employing physical neural architectures and elements of artificial intelligence in the mechanical domain have arguably gained more traction \cite{lee2022mechanical}. Ranging from binary-stiffness beams which learn desired shape-morphing behaviors \cite{hopkins2023using}, to Hopf oscillators \cite{shougat2021hopf, shougat2022dynamic}\Revision{, phononic \cite{zhang2023embodying} and multi-stable metamaterials \cite{liu2023cellular}}, there has been a spurt of activity demonstrating mechanical learning via programmable structures at different length scales. \Revision{Additionally, recent investigations have demonstrated the utility of origami structures in physical reservoir computing, leveraging the multistability and reconfigurability of the Muira-ori folding technique \cite{Liu2023,wang2023building}, and exploiting the nonlinear folding dynamics of dynamic truss-frame models \cite{bhovad2021physical}, to perform intelligent tasks such as input recognition, classification, and even robotic-related functions such as emulation and pattern generation.} At their very core, neuromorphic computers (initially proposed by Mead \cite{mead1989analog, mead1990neuromorphic}) seek to mimic the intricate workings of the brain's nervous system and its vastly distributed nature \cite{sarkar2022organic, gkoupidenis2022artificial,van2018organic,wright2022deep,prucnal2017neuromorphic,shastri2021photonics, merolla2014million}. There are two main approaches to develop neuromorphic systems. The first involves transferring and translating existing digital neural architectures to physical substrates, while the second pertains to creating new algorithms which more accurately emulate the functionality of biological neurons and synapses \cite{markovic2020physics}. \Revision{At the hardware level, the implementation of neuromorphic computing has traditionally been achieved using various techniques ranging from diffractive optics \cite{lin2018all,zhou2021large, leonard2022high} to integrated photonics \cite{shen2017deep, ballarini2020polaritonic}, memristors \cite{li2018review, xia2019memristive}, and MEMS-based neural networks \cite{barazani2020microfabricated,nikfarjam2023energy}}. Most recently, neuromorphic metasurfaces have emerged as a novel approach to achieve neural functions in optical \cite{fu2023photonic, leonard2021co, wu2020neuromorphic} and acoustic \cite{weng2020meta} forms. In these systems, the neuromorphic inference is rooted in the trained wave behavior of a set of metasurfaces comprised of tunable unit cells (physical neurons). These metasurfaces, spaced along the direction of wave propagation, form the basis of a neural network, and the subsequent wave scattering of a physical agent (e.g., light or sound) in each of the physical layers constitutes the interactions within these layers. A readout mechanism is then used to interpret the results culminating in a desired output. 

While recent advances have been made in the development of photonic neuromorphic metasurfaces, much less has been accomplished in the realm of mechanics. In here, we present a first attempt to describe the fundamentals of an elastic neuromorphic metasurface, which instills a level of cognitive intelligence in a multi-layered classifier, enabling it to mechanically execute neural network classification tasks while overcoming a new set of challenges associated with the complex nature of elastic waves with respect to scattering and manipulation. Despite the understandable trade-off in computational pace and precision associated with elastoacoustic waves (compared to their optical counterparts), the goal here is not to compete with the performance metrics achieved via photonic systems, but rather provide a minimal threshold of intelligence and computational function in core mechanical parts and metallic structures which already constitute a core component of several equipment and devices. This form of mechanical computing is even more prudent in applications where the computational input is readily available in mechanical form (e.g., vibrations, dynamic loads, noise, or impinging acoustic waves) thus eliminating the need for encoding the input features, or extreme environments which render electronic parts dysfunctional (e.g., high magnetic fields or elevated temperatures). 

In addition to the aforementioned challenges, we seek to resolve another limitation of wave-based neuromorphic systems, namely, their inability to be reconfigured to conduct a new task, differing from what it was initially designed and trained for, unless the system is fully dismantled and/or reconstructed. In this work, we demonstrate the inner-workings of multiple layers of reconfigurable waveguides, which are phase-trained via constant weights and trainable activation functions in a manner that enables the resultant wave scattering at the readout location to focus on the correct class label within the detection plane. A comprehensive framework detailing the theory behind the unit cell, waveguide array, and overall metasurface designs is provided, and is then used to reconstruct the classic MNIST problem using custom tailored hyperparameters of the proposed system. Finally, shifting from the MNIST to the Iris dataset, we demonstrate the neuromorphic system's ability to achieve full reconfigurability and perform distinct classification tasks without the need for remanufacturing.

%%%%%%%%%%%%%%%%%%%%%%%%%%%%%%%%%%%%%%%%%%%%%%%%%%%%%%%%%%%%%%%%%%%%%%%%%%%%%%%%%%%%%%%%%%%%%%
%%%%%%
\begin{figure*}[t!]
\centering
\includegraphics[width=\textwidth]{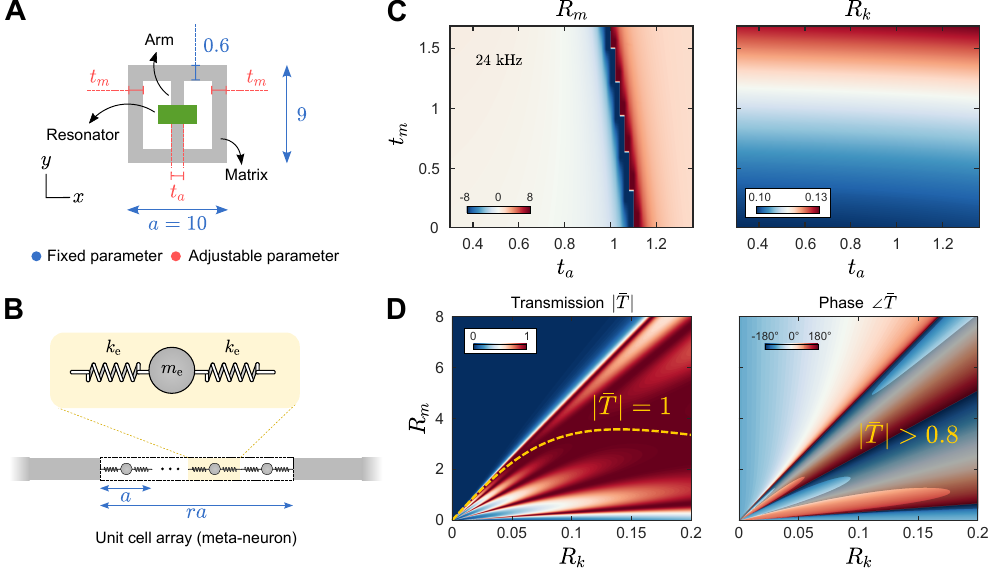}
\caption{\textbf{Waveguide design and transmission profile.} (A) Reconfigurable unit cell realized as a solid continuum. Blue and red dimensions denote fixed and adjustable parameters. (B) Equivalent lumped parameter representation of the unit cell. $m_e$ and $k_e$ are the effective mass and stiffness, respectively, and $a$ is the lattice constant. (C) Effective mass and stiffness ratios, $R_m$ and $R_k$, as functions of $t_a$ and $t_m$ at $24$ kHz. (D) Transmission amplitude $|\bar{T}|$ and phase $\angle \bar{T}$ for a waveguide comprised of $r=6$ unit cells. The dashed line in $|\bar{T}|$ indicates a full transmission scenario ($|\bar{T}| = 1$), while the mask in $\angle \bar{T}$ represents $|\bar{T}| > 0.8$ regions. Schematic diagrams are not drawn to scale and all dimensions are in mm.}
\label{Fig1}
\end{figure*}
%%%%%%
\section{\label{sec:neuromorphic_metasurface}NEUROMORPHIC METASURFACE}
\subsection{\label{sec:theory}Concept}
We present an elastic neuromorphic metasurface composed of 3 sections: input gates, metasurface neurons, and detection units. These are equivalent to the input, hidden, and output layers of a digital neural network, respectively. The system can be excited at the input gates via any excitational signal which can be encoded into a vibrational waveform (e.g., noise, image, or dynamic load), culminating in elastic waves which propagate through the medium encompassing the metasurface layers. In current applications, the phase profile of a metasurface can be designed using an appropriate delay law to focus or steer a wavefront in a given direction. In this work, however, we will show the ability of a neuromorphic metasurface to generate a desirable output from a given input via intelligent, interlayer wave scattering. Provided with a set of tunable degrees of freedom (in this case, phase gradients between neighboring metasurface cells), the neuromorphic system trains itself to attain an intricate configuration of \textit{meta-neurons} which allow the fully-assembled elastic structure to execute the required task.

While the model shown here takes inspiration from digital neural networks, its architecture is notably different in important ways, in order to accommodate the nuanced mechanics of elastic wave propagation. A conventional digital neural network is comprised of trainable weights and biases between the network layers. However, the activation function applied at each layer is typically non-trainable. In a neuromorphic metasurface, we will show that such weights represent wave scattering characteristics which are constrained functions of the chosen geometry, inertial, and stiffness properties. Once specified, these weights remain unchanged and are, therefore, non-trainable. Instead, phase delays within the metasurface layers provide an alternative tunable (and thus trainable) platform, shown to be analogous to a trainable activation function in a neural network framework.
%%%%%%%%%%%%%%%%%%%%%%%%%%%%%%%%%%%%%%%%%%%%%%%%%%%%%%%%%%%%%%%%%%%%%%%%%%%%%%%%%%%%%%%%%%%%%%
%%%%%%
\begin{figure*}[t!]
\centering
\includegraphics[width=\textwidth]{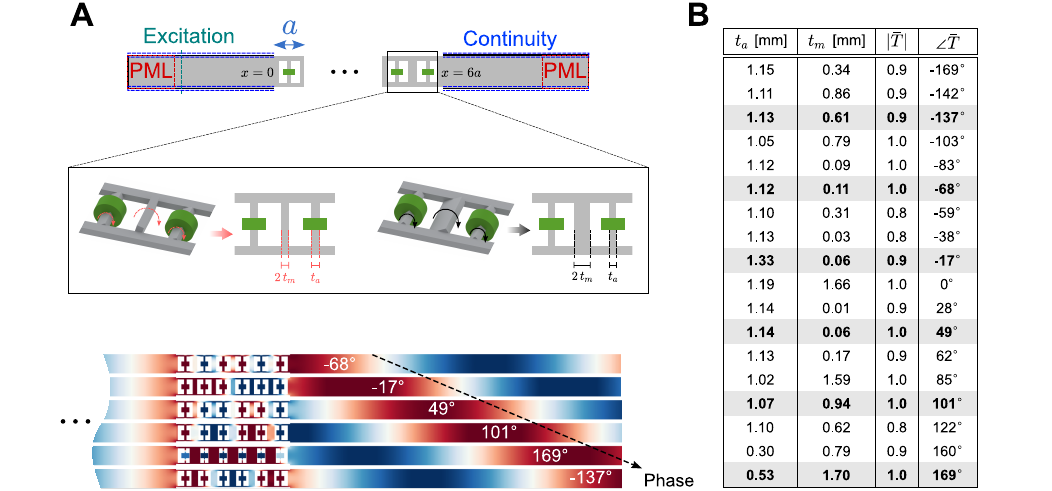}
\caption{\textbf{Waveguide design.} (A) Displacement field for the selected waveguides highlighted using a dark background in (B), obtained via finite element modeling. The top panel depicts the details of the numerical simulations showing locations of the continuity boundary conditions and the perfectly matched layer (PML). Shown in the middle is a 3D physical realization of the reconfigurable waveguide design, exploiting the change in bending stiffness for rotating cross sections to achieve the desired tunability. (B) Waveguides designed using $R_{k,m}$ and $\bar{T}$ maps in order to exhibit high transmission at $20^\circ$ phase differences over a full $360^\circ$ range.}
\label{Fig2}
\end{figure*}
%%%%%%
\subsection{\label{sec:Unitcell}Unit cell}
Driven by the need for a mechanically-tunable system, we utilize a reconfigurable locally resonant unit cell which exhibits subwavelength scattering. Since computational pace is directly proportional to the wave speed in the medium, the focus here is on axial vibrations of the resonator given the higher speed of longitudinal wave propagation compared to flexural waves. The solid unit cell shown in Fig.~\ref{Fig1}A is designed to have effective properties, $m_e$ and $k_e$, equivalent to these of the lumped parameter representation (depicted in Fig.~\ref{Fig1}B), which can be retrieved from finite element simulations of the solid cell via transfer matrix equivalence (see SI Appendix Note 1). The unit cell shown in Fig.~\ref{Fig1}A is composed of an Aluminum matrix ($E=70$ GPa, $\rho=2700$ kg/m$^3$ and $\nu=0.33$), a Brass resonator ($E=97$ GPa, $\rho=8490$ kg/m$^3$ and $\nu=0.31$) and two slender Aluminum beams (resonator arms) which attach the resonator to the matrix. These arms are placed along the $y$-axis, perpendicular to the longitudinal deformation of the unit cell along the $x$-direction, enabling a coupling between the bending modes of the resonator arms and the axial vibrations of the unit cell.

At the core of the neuromorphic metasurface theory is the ability to achieve full phase tunability over a $2\pi$ range while maintaining high transmission within the waveguide. The unit cell, therefore, needs to be reconfigurable in order to admit different values of $m_e$ and $k_e$ as needed. The effective widths of the resonator arm $t_a$ and the vertical section of the unit cell matrix $t_m$, both indicated on Fig.~\ref{Fig1}A, play a central role in the bending stiffness of both parts. Thus, changes in these two parameters significantly alter the dynamics of the unit cell described by $m_e$ and $k_e$. Control over these effective widths can be exercised via a 3D realization of the unit cell with rotating arms having rectangular cross-sections, as shown in Fig.~\ref{Fig2}A and successfully implemented in literature \cite{attarzadeh2020experimental}. We define the effective mass and stiffness ratios as $R_m = m_e / m_{\text{slab}}$ and $R_k = k_e / 2 k_{\text{slab}}$, respectively, where $m_{\text{slab}} = \rho A a$ and $k_{\text{slab}} = E A / [a(1-\nu^2)]$ denote the mass and axial stiffness, respectively, of an aluminum slab of the same size. Figure~\ref{Fig1}C depicts the variation of $R_m$ and $R_k$ as functions of $t_a$ and $t_m$ at a frequency of $24$ kHz. The behavior shows that while both thicknesses affect both effective ratios (i.e., mass and stiffness), $R_m$ and $R_k$ are observed to be largely dependent on $t_a$ and $t_m$, respectively.
%%%%%%%%%%%%%%%%%%%%%%%%%%%%%%%%%%%%%%%%%%%%%%%%%%%%%%%%%%%%%%%%%%%%%%%%%%%%%%%%%%%%%%%%%%%%%%
%%%%%
\begin{figure*}[t!]
\centering
\includegraphics[width=\textwidth]{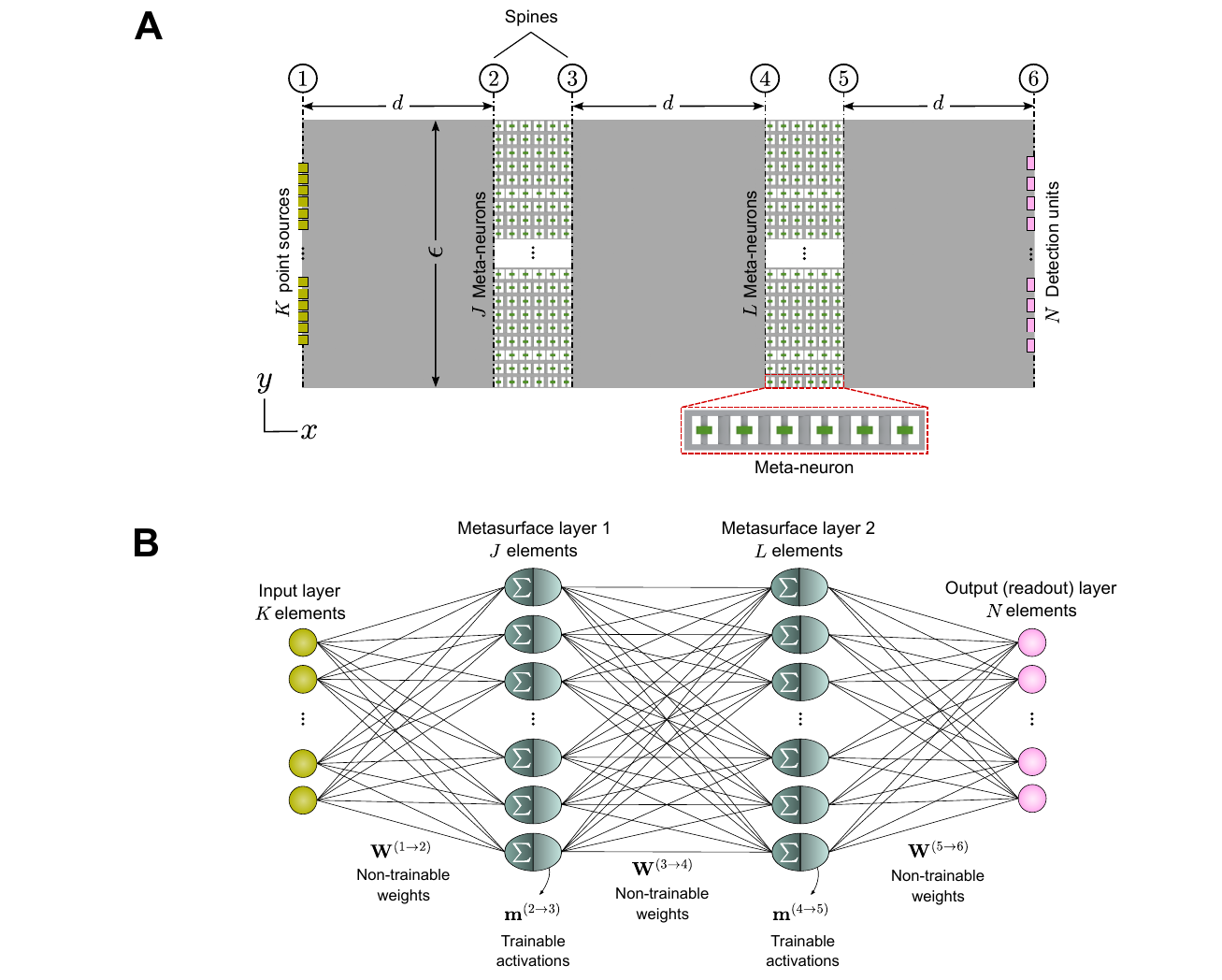}
\caption{\textbf{Reconfigurable elastic neuromorphic metasurface.} (A) Schematic illustration of the assembled neuromorphic system. Six vertical cross sections, referred to as spines and labeled $1$ through $6$, define the different mechanical components. Spines $1$ and $6$ mark the input and output (readout) layers, with $K$ sources and $N$ detection units. Two metasurface layers span the regions between spines $2 \to 3$ and $4 \to 5$ housing $J$ and $L$ meta-neurons, respectively. Each meta-neuron consists of $6$ identical unit cells and retains a unique design based on tuned values of $t_a$ and $t_m$ (see Fig.~\ref{Fig1}). Regions before, between, and after the metasurface layers are homogenous aluminum plates. (B) Logic diagram depicting the neural architecture of the elastic neuromorphic system, emphasizing the non-trainable weights $\mathbf{W}^{(1\rightarrow 2)}$, $\mathbf{W}^{(3\rightarrow 4)}$, and $\mathbf{W}^{(5\rightarrow 6)}$, as well as the trainable activation functions $\mathbf{m}^{(2\rightarrow 3)}$ and $\mathbf{m}^{(4\rightarrow 5)}$. All layers are dense.}
\label{Fig3}
\end{figure*}
%%%%%
\subsection{\label{sec:Waveguide}Scattering and waveguide design}
The scattering properties of a waveguide comprising $r$ unit cells can be evaluated starting with the dispersion relation, which can be derived as follows~\cite{lee2018mass}:
\begin{equation}
\label{eq1}
    m_e \omega^2 = 2 k_e \sin^{2}(\frac{\kappa a}{2})
\end{equation}
where $\kappa=\omega/c$ is the Bloch wavenumber and $\omega$ is the angular frequency of the harmonic wave. The transmission coefficient $\bar{T}$ can be computed from:
\begin{equation}
\label{eq2}
    \bar{T} =  \frac{2}{2 \cos{(r \kappa a)} + \ii (z z_s^{-1} + z_s z^{-1}) \sin{(r \kappa a)}}
\end{equation}
where $\ii$ is the imaginary unit, $z = \rho A c$ is the mechanical impedance, and $c = \sqrt{ E / [\rho(1-\nu^2)]}$ is the axial wave speed (See SI Appendix Note 2 for detailed derivation). The system's characteristic impedance $z_s$ can be defined as the ratio of $\sqrt{\frac{m_e k_e}{2}}$ to $\cos{(\frac{\kappa a }{2})}$. For a waveguide consisting of $r=6$ unit cells, shown in the bottom panel of Fig.~\ref{Fig1}B, we start by considering a wide range of $m_e$ and $k_e$ values, and obtain $\kappa$ from Eq.~(\ref{eq1}). This is then used to determine the corresponding $z_s$, allowing the transmission amplitude $|\bar{T}|$ and phase $\angle \bar{T}$ to be evaluated by using Eq.~(\ref{eq2}), as shown in Fig.~\ref{Fig1}D. A high transmission ($|\bar{T}|>0.8$) mask is applied to the phase map to facilitate the selection of $R_k$ and $R_m$ combinations that maintain a minimum acceptable transmission at a given phase.

As indicated earlier, the neuromorphic metasurface needs to maintain sufficient tunability for good training and subsequent performance. Making use of $R_{k,m}$ (Fig.~\ref{Fig1}C) and the transmission amplitude and phase maps (Fig.~\ref{Fig1}D), eighteen arrays are generated with intervals of approximately $20^\circ$ phase difference as summarized in the table provided in Fig.~\ref{Fig2}B. These can be the building blocks of the neuromorphic metasurface.
Finite element modeling is used to validate the obtained transmission and phase delays for six select cases from the table, and the resultant displacement field is demonstrated in Fig.~\ref{Fig2}A.
Finally, in the culminating neuromorphic metasurface, each unit cell array in the metasurface is treated as a singular point, neglecting lateral propagating waves. As such, the Aluminum slabs are chosen to be $1$ mm thicker than the array itself to mitigate vertical interactions that are unaccounted for and ensure isolation from adjacent arrays.
%%%%%%%%%%%%%%%%%%%%%%%%%%%%%%%%%%%%%%%%%%%%%%%%%%%%%%%%%%%%%%%%%%%%%%%%%%%%%%%%%%%%%%%%%%%%%%
\section{\label{sec:operational_theory}OPERATIONAL THEORY}
\subsection{\label{sec:Neural_architecture}Neural architecture}
The neuromorphic metasurface's neural architecture consists of four consecutive layers: an input layer, two metasurface layers, and an output layer, as illustrated in Fig.~\ref{Fig3}.
These are equivalent to the input, hidden, and output layers of a digital neural network, respectively.
For a given classification problem, training and test samples with $K$ features are reshaped into a single dimensional vector with the same number of elements. This vector is then fed to the elastic medium in the form of $K$ displacement point sources at the input layer, with the input features manifesting themselves in the amplitude of each excitation point. At the far side of the neuromorphic metasurface (output layer), a set of $N$ detection units are defined corresponding to the $N$ labels of the classification problem, as a computational readout. Each waveguide is comprised of an array of $r=6$ unit cells, henceforth referred to as a ``meta-neuron''. For a successful mechanical neural network, the goal of the training is to distinctly engineer the two metasurface layers such that for a given input (i.e., test sample) generating a distinct excitation pattern stemming from the $K$ point sources constituting this input, the scattered elastic wavefronts would predominantly focus on the appropriate detection unit spatially corresponding to the correct $N$ label.
While the model shown here takes inspiration from digital neural networks as can be seen in Fig.~\ref{Fig3}B, its architecture is notably different in important ways, in order to accommodate the nuanced mechanics of elastic wave propagation. A conventional digital neural network is comprised of trainable weights and biases between the network layers. However, the activation function applied at each layer is typically non-trainable. In a neuromorphic metasurface, we show that such weights represent wave scattering characteristics which are constrained functions of the chosen geometry, inertial, and stiffness properties. Once specified, these weights remain unchanged and are, therefore, non-trainable. Instead, phase delays within the metasurface layers provide an alternative tunable (and thus trainable) platform, shown to be analogous to a trainable activation function in a neural network framework. The analytical framework, detailed in SI Appendix Note 3, tracks the trajectory of incident waves as they propagate through the different components, evaluating wave amplitudes at the six spines labeled $1$ through $6$ in Fig.~\ref{Fig3}A.

A conventional neural network is mathematically described by:
\begin{equation}
    \mathbf{y}^{(i+1)}=f(\mathbf{b}^{(i+1)}+\mathbf{W}^{(i\rightarrow i+1)} \mathbf{y}^{(i)})
\end{equation}
where $\mathbf{y}^{(i)}$ is a vector of values assigned to neurons in the $i^{\text{th}}$ layer, $\mathbf{W}^{(i\rightarrow i+1)}$ is the weights matrix, and $\mathbf{b}^{(i+1)}$ is the bias. The architecture of the elastic neuromorphic system is similar to the above framework in that $\mathbf{y}^{(i)}$ is analogous to the displacement vector for the $i^{\text{th}}$ layer $\mathbf{u}^{(i)}$ and the weights matrix $\mathbf{W}^{(i\rightarrow i+1)}$ is equivalent to the wave propagation matrix from one spine to the next. Furthermore, the activation function $f$ is analogous to the metasurface effect inflicted by $\mathbf{m}^{(i\rightarrow i+1)}$ vector on the incident wave. However, the bias $\mathbf{b}^{(i+1)}$ is not applicable in this study. More importantly, in the elastic neuromorphic system, the wave propagation matrices do not change once dimensions and material properties are finalized, rendering them \textit{hyperparameters}, contrary to the weights of a digital neural network which are the primary trainable parameters throughout the learning process. Another stark contrast is the activation function applied at each layer of a digital neural network, which is typically identical for all neurons. In the elastic neuromorphic system, however, phase shifts applied by each meta-neuron of the metasurface layers (and consequently the $\mathbf{m}^{(i\rightarrow i+1)}$ vectors) are trainable parameters in the learning algorithm. The neuromorphic metasurface's wave manipulation approach can be perceived as multiple beam-forming segments combined together in one line and steering separate parts of the incident wave.
%%%%%%%%%%%%%%%%%%%%%%%%%%%%%%%%%%%%%%%%%%%%%%%%%%%%%%%%%%%%%%%%%%%%%%%%%%%%%%%%%%%%%%%%%%%%%%
\subsection{\label{sec:MNIST_acoustic}Encoding of input data features}

To illustrate the performance of the elastic neuromorphic metasurface, and without loss of generality, we utilize the classical MNIST dataset which is widely used for training and testing in the field of machine learning. While the system presented here is ideally suited for inputs which take a mechanical form, MNIST has served as a commonly-adopted benchmark in the context of photonic \cite{wu2020neuromorphic} and even acoustic \cite{weng2020meta} neuromorphic metasurfaces in a manner that goes beyond optical digit recognition, owing to its simplicity and high diversity. As will be detailed later, the MNIST data used here is eventually fed into the neuromorphic metasurface merely as a set of vibrational excitations of varying amplitude and, as such, are taken at their face values (raw numbers) which can be interpreted differently for different applications.

\begin{figure*}[t!]
\centering
\includegraphics[width=\textwidth]{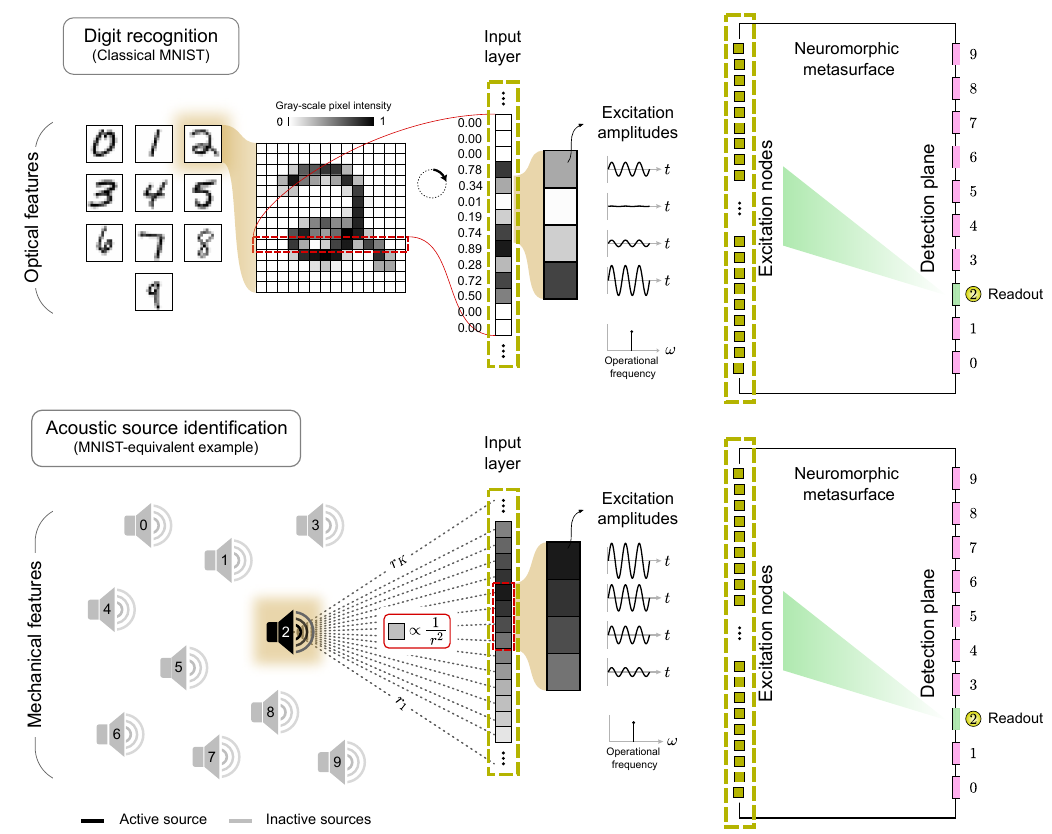}
\caption{\textbf{Encoding of input data.} The top row shows the pre-processing of a single digit sample ($2$) from the MNIST dataset prior to being fed into the neuromorphic metasurface. The bottom row depicts a different interpretation of the same problem where the elastic deformations at the input excitation nodes correspond to pressure amplitudes reaching each input node as a result of an active sound-emanating source which is a given distance away, following the acoustic inverse square law. The output layers in both examples consist of 10 detection units that are perceived as either digits or speakers, corresponding to the possible classes of the classification problem.}
\label{Fig4}
\end{figure*}

Each sample in the MNIST dataset consists of a $28\times28$ matrix of gray-scale pixel intensities. The matrix is flattened into a 1D vector and the pixel intensities are used as displacement amplitudes for $K = 784$ sinusoidal excitation point sources at the input layer which share the same operational frequency $\omega$, as depicted in the top panel of Fig.~\ref{Fig4}. At the far end of the neuromorphic metasurface (output layer), $N=10$ detection units are defined corresponding to the possible input classifications: $0$ to $9$. A properly trained and designed metasurface is capable of focusing most of the energy in the detection layer at the correct output unit, indicating the recognized digit as the classifier's readout. The bottom panel of the figure depicts a different interpretation of the same problem where the elastic deformations at the $K$ locations correspond to pressure amplitudes reaching each input node as a result of an active sound-emanating source which is a euclidean distance $r_i$, $1 \leq i \leq K$, away. This equivalency between the two tasks demonstrates the system's comparable performance for both the pre-processed MNIST and acoustic tasks, and the system's ability to leverage mechanical environments to carry out a meaningful computational task on an input which already exists in the system's native physical domain.

%%%%%%
%%%%%%
%%%%%%%%%%%%%%%%%%%%%%%%%%%%%%%%%%%%%%%%%%%%%%%%%%%%%%%%%%%%%%%%%%%%%%%%%%%%%%%%%%%%%%%%%%%%%%
\subsection{\label{sec:training}Training}

The metasurface layers are set to be $d=2$~m apart with $J=L=1,000$ meta-neurons in each metasurface layer. The distance between neighboring point sources at the input layer is set to $10$~mm, equal to the width of a single meta-neuron (see Section~\ref{sec:hyper} for an optimization of geometrical and hyperparameters). The model is developed in TensorFlow using custom Keras layers and is trained in Python. The flattened input samples are cast into a vector of complex numbers $\mathbf{u}^{(1)}_{784\times 1}$ since the weights and the phase shift have to be complex in order to retain the displacement components from one layer to another. Two custom dense layers are then defined in the model for the two metasurface layers, with non-trainable weights ($\mathbf{W}^{(1\rightarrow 2)}_{1000\times 784}$ and $\mathbf{W}^{(3\rightarrow 4)}_{1000\times 1000}$ from Eqs.~(S23) and (S27), respectively), no biases, and custom activation functions ($\mathbf{m}^{(2\rightarrow 3)}_{1000\times 1}$ and $\mathbf{m}^{(4\rightarrow 5)}_{1000\times 1}$ from Eqs.~(S24) and (S28), respectively). As detailed in SI Appendix Note 4, the codependency between the transmission amplitude and the phase delay in each meta-neuron is considered when applying the activation functions in the training process. Lastly, an additional custom dense layer is defined as the output layer with non-trainable weights ($\mathbf{W}^{(5\rightarrow 6)}_{1000\times 1000}$) and with no biases. In this final layer, the well-known Softmax activation function is implemented.

While the inference step of the neuromorphic metasurface is purely mechanical, the training takes place in a computer \cite{markovic2020physics} equipped with an Intel Xeon\textsuperscript{\tiny\textregistered} Gold 6230 CPU @ $2.10$ GHz and a RAM of 128 GB, running on Microsoft Windows 10. The learning is carried out using $60,000$ training samples and a $10\%$ split for validation. The Sparse Categorical Cross-entropy loss function, which is highly effective in multi-class classification problems, is adopted, and the Adam optimizer is employed to improve the training performance via a defined learning rate schedule. After training, the model is evaluated using $10,000$ (blind) test samples to report its accuracy. Upon ensuring that the training has been executed with high efficacy, the trained metasurface vectors $\mathbf{m}^{(i\rightarrow i+1)}$ are used to determine the corresponding design parameters $t_a$ and $t_k$ for each and every meta-neuron.
%%%%%%%%%%%%%%%%%%%%%%%%%%%%%%%%%%%%%%%%%%%%%%%%%%%%%%%%%%%%%%%%%%%%%%%%%%%%%%%%%%%%%%%%%%%%%%
%%%%%%
\begin{figure*}[t!]
\centering
\includegraphics[width=\textwidth]{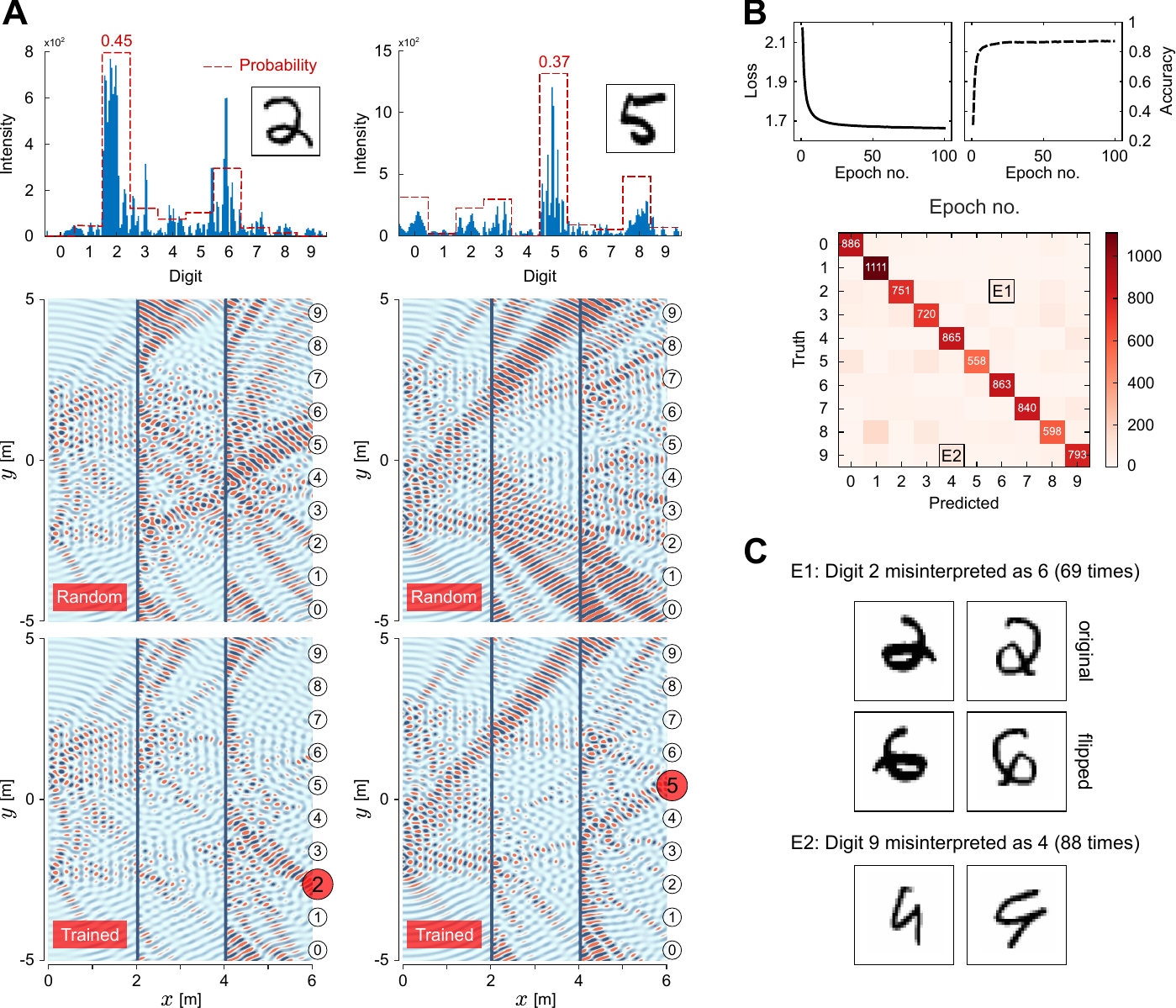}
\caption{\textbf{Performance of the elastic neuromorphic metasurface.} (A) Output of the neuromorphic metasurface for two digit recognition test samples from the MNIST dataset corresponding to the digits ``$2$'' (left) and ``$5$'' (right). The top row shows the intensities (blue bars) and probabilities (red dashed lines) at different detection units within the output (readout) layer. Bottom and middle rows show the scattered wavefields for trained and randomly initialized neuromorphic metasurfaces, respectively. (B) Classification accuracy (right) and model loss (left) for the training dataset, and confusion matrix (bottom) for the testing dataset. The training dataset is divided into 30 batches processed at each epoch. (C) Examples of erroneous recognition: ``$2$'' and ``$9$'' misinterpreted as ``$6$'' and ``$4$'', respectively, due to subtle shape similarities, corresponding to the E1 and E2 markings on the confusion matrix in (B).}
\label{Fig5}
\end{figure*}
%%%%%%
\section{\label{sec:performance}PERFORMANCE}
Figure~\ref{Fig5}A showcases the output of the neuromorphic metasurface for two digit recognition samples corresponding to the digits ``$2$'' and ``$5$''. The top row illustrates the distribution of intensity and probability over the output layer at different detection units. The latter is computed as the ratio of the summation of point intensities at the detection units to the total intensity of the output layer. As evident from these plots, the neuromorphic metasurface is able to successfully classify both samples. The shown wavefields at the bottom of Fig.~\ref{Fig5}A further corroborate this by depicting the wave scattering through the different layers of the neuromorphic metasurface from the input location at spine $1$ to the output (readout) location at spine $6$. The displacements in each of the three sections shown ($1 \to 2$, $3 \to 4$, and $5 \to 6$) are individually normalized with respect to the same section for better visualization. The middle row of Fig.~\ref{Fig5}A displays the wavefield resulting from a randomized, untrained design, further confirming the effectiveness of the trained wave scattering.

The top panel of Fig.~\ref{Fig5}B summarizes the loss and accuracy as functions of the epoch number. An accuracy of $87.5$\% in $100$ epochs was achieved taking approximately $2.6$~hours of training on the reported PC in Section~\ref{sec:training}. Wavefields showing successful detection of all numeric digits are provided in SI Appendix Note 5. The overall performance is evaluated using a confusion matrix for the trained model based on its performance on the entire testing dataset, as demonstrated in the bottom panel of Fig.~\ref{Fig5}B. While this provides important insights into the model's capability of correctly classifying the tested samples, it also helps identify potential areas of improvement, such as recognizing specific digit shapes that may be challenging to accurately predict. As a case in point, the labels E1 and E2 marked on the confusion matrix correspond to erroneous recognition of the digits ``$2$'' and ``$9$'' as $6$ and $4$, respectively, due to subtle shape similarities illustrated in Fig.~\ref{Fig5}C.
%%%%%%%%%%%%%%%%%%%%%%%%%%%%%%%%%%%%%%%%%%%%%%%%%%%%%%%%%%%%%%%%%%%%%%%%%%%%%%%%%%%%%%%%%%%%%%
\section{\label{sec:hyper}HYPERPARAMETERS}

The size and dimensions of the elastic neuromorphic metasurface are scalable, and can be downsized depending on the operational frequency and waveguide design. Moreover, the computational accuracy of the elastic neuromorphic metasurface is affected by several hyperparameters that, if adequately tuned, can considerably improve several performance metrics. These are aspects which are set at the design stage and control the learning process, as opposed to being derived via training. The contribution of each of these to the overall classification accuracy is evaluated by individually altering parameters of interest while holding the rest unchanged from the values reported earlier, as shown in Table~\ref{Table1}. It is important to acknowledge that, for efficiency purposes, the model used in these studies is trained for a predetermined number of epochs, reaching a near-stabilized accuracy level, while the training duration may be shorter compared to the main model.

\begin{table}[t!]
\centering
\caption{Effect of different hyperparameters stated in the left column of the table on the overall classification accuracy. Cases typed in boldface refer to the main model.}
\begin{tabular}{ccccc} \toprule

\multirow{2}{*}{\begin{tabular}{c}No. of layers\end{tabular}} & \textbf{2} & 3 & 4 & 5\\ \cmidrule(l{3pt}r{3pt}){2-2} \cmidrule(l{3pt}r{3pt}){3-3} \cmidrule(l{3pt}r{3pt}){4-4} \cmidrule(l{3pt}r{3pt}){5-5}
 & \textbf{87.5\%} & 90.7 & 91.9\% & 93.0\% \\ \midrule
 
\multirow{2}{*}{\begin{tabular}{c}Distance\\between layers\end{tabular}} & 0.3 m & 1 m & \textbf{2 m} & 3 m \\ \cmidrule(l{3pt}r{3pt}){2-2} \cmidrule(l{3pt}r{3pt}){3-3} \cmidrule(l{3pt}r{3pt}){4-4} \cmidrule(l{3pt}r{3pt}){5-5}
 & 87.0\% & 87.3\% & \textbf{87.5\%} & 86.9\% \\ \midrule

\multirow{2}{*}{\begin{tabular}{c}No. of neurons\end{tabular}} & 100 & 250 & 784 & \textbf{1,000} \\ \cmidrule(l{3pt}r{3pt}){2-2} \cmidrule(l{3pt}r{3pt}){3-3} \cmidrule(l{3pt}r{3pt}){4-4} \cmidrule(l{3pt}r{3pt}){5-5}
 & 45.1\% & 70.7\% & 86.7\% & \textbf{87.5\%} \\ \midrule

\multirow{2}{*}{\begin{tabular}{c}Width of \\ detection units \end{tabular}} & 0.3 m & 0.5 m & 0.7 m & \textbf{1 m} \\ \cmidrule(l{3pt}r{3pt}){2-2} \cmidrule(l{3pt}r{3pt}){3-3} \cmidrule(l{3pt}r{3pt}){4-4} \cmidrule(l{3pt}r{3pt}){5-5}
 & 89.2\% & 88.7\% & 88.3\% & \textbf{87.5\%} \\ \midrule
 
 \multirow{2}{*}{\begin{tabular}{c}Training metric\end{tabular}} & \multicolumn{2}{c}{\textbf{Intensity}} & \multicolumn{2}{c}{Displacement}\\ \cmidrule(l{6pt}r{6pt}){2-3} \cmidrule(l{6pt}r{6pt}){4-5}
 & \multicolumn{2}{c}{\textbf{87.5\%}} & \multicolumn{2}{c}{82.5\%} \\ \midrule
 
 \multirow{2}{*}{\begin{tabular}{c}Digit processing\end{tabular}} & \multicolumn{2}{c}{\textbf{Gray-scale}} & \multicolumn{2}{c}{Binary}\\ \cmidrule(l{6pt}r{6pt}){2-3} \cmidrule(l{6pt}r{6pt}){4-5}
 & \multicolumn{2}{c}{\textbf{87.5\%}} & \multicolumn{2}{c}{85.8\%} \\ \bottomrule
\end{tabular}
\label{Table1}
\end{table}

\Revision{As can be inferred from the first row of Table~\ref{Table1}, increasing the number of metasurface layers improves the accuracy in a proportional manner (See SI Appendix Note 6 for details). A similar trend is observed for both the distance between layers $d$ and the number of neurons in each layer ($J$ or $L$). Nonetheless, a trade-off between accuracy and practicality is unavoidable. Accuracy gains become insignificant beyond a certain range while the structural size multiplies, detrimentally influencing wave intensity at the detection units, which, in turn, adversely affects the accuracy as noted in the last column. Contrarily, reducing the dimensions of the detection units marginally enhances the accuracy which, while appearing advantageous, concurrently permits a substantial portion of the energy to be redirected away, resulting in low intensities at the detection units. As evident from the fifth row of Table~\ref{Table1}, employing the intensity of the wavefield at the output layer as the evaluation metric yields improved accuracy compared to the displacement. Finally, although the utilization of binary simplifications for the sample digit pixels at the input layer could offer more flexibility in fabrication and testing, allowing on/off actuation, reducing the number of sources through combination, and eliminating the need for amplitude adjustment at the excitation location, it is observed that a gray-scale representation leads to a slightly improved accuracy, as can be seen in the last row of Table~\ref{Table1}.}

\Revision{Beyond the discussed hyperparameters, another important factor is the system's scalability, which directly influences practical implementation. This includes the total number of meta-neurons and the overall footprint (length and width) affected by the chosen number of layers, the number of meta-neurons per layer, and the inter-layer spacing. To explore this aspect, we re-engineered the system for significant downsizing using two design variations. Both designs utilize five layers, with either 250 or 100 meta-neurons per layer, and achieve reductions of $92.5\%$ and $98\%$, respectively of the overall area (see SI Appendix Note 7). While significantly smaller in size and meta-neuron count compared to the original model, the downsized designs achieved accuracy levels of $79\%$ and $75\%$, respectively. While this confirms the inherent trade-off between accuracy and practicality, it is clear that the system maintains an acceptable performance even when substantially reduced in scale.} 

It is worth noting that in addition to the optimization of the aforementioned parameters for enhanced accuracy and an overall better performance, there exists the possibility of training the homogenous plate domains before and after the metasurface layers. This could be accomplished through the utilization of graded stiffness or thickness profiles or optimized topological configurations. While these domains would not be re-trainable like the reconfigurable metasurface layers, they could conceivably be fine-tuned to sit around a handful of input-output datasets, such that the active domains can efficiently map each of the individual tasks.
%%%%%%%%%%%%%%%%%%%%%%%%%%%%%%%%%%%%%%%%%%%%%%%%%%%%%%%%%%%%%%%%%%%%%%%%%%%%%%%%%%%%%%%%%%%%%%
\section{\label{sec:reconfig}RECONFIGURABILITY}
%%%%%%
\begin{figure*}[t!]
\centering
\includegraphics[width=\textwidth]{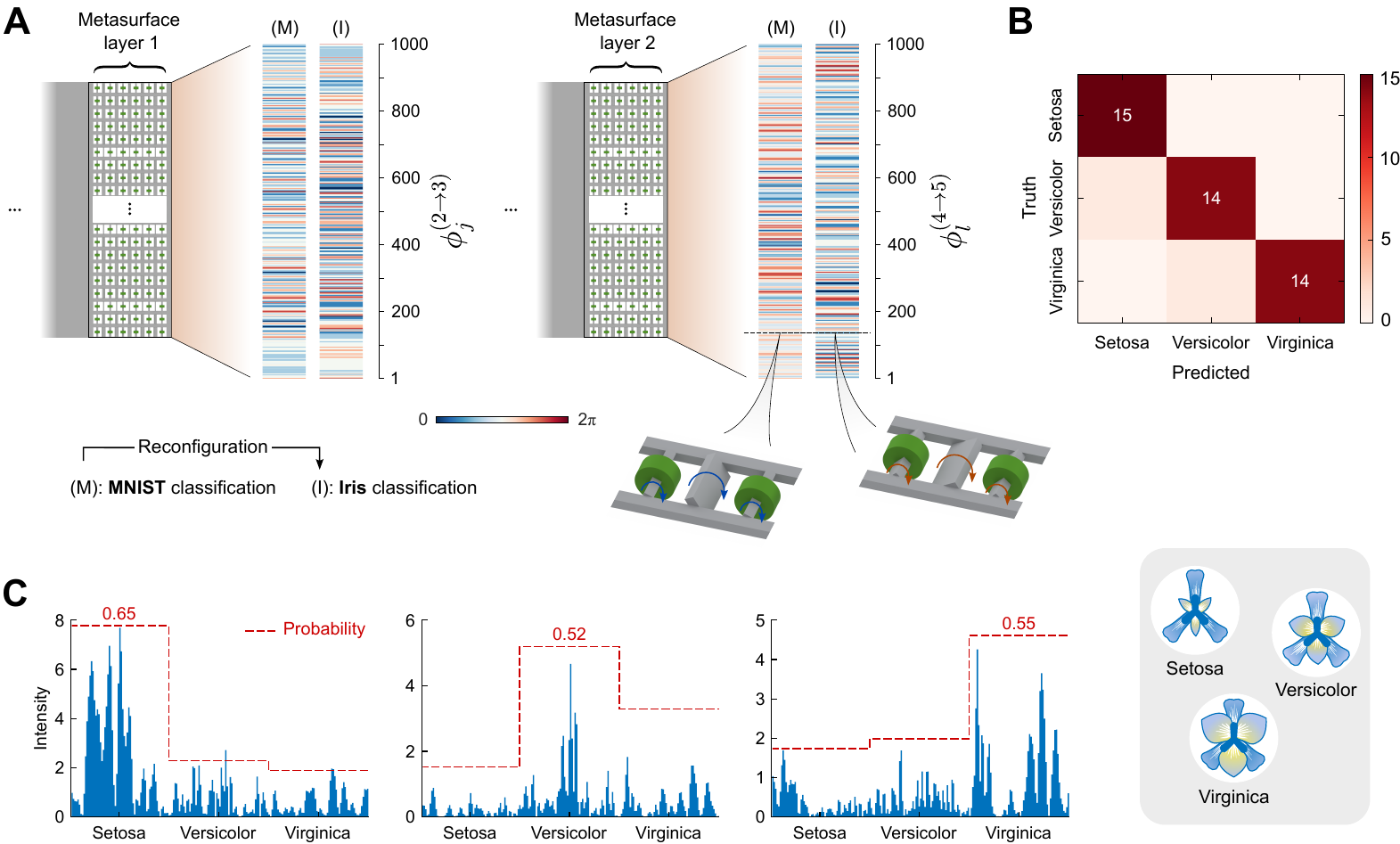}
\caption{\textbf{Reconfigurability.} (A) Change in the phase profile of the 1,000 meta-neurons in the first (left) and second (right) metasurface layers following training as the neuromorphic system switches from the MNIST (M) to the Iris (I) classification task. $\phi_{j}^{(2\rightarrow3)}$ and $\phi_{l}^{(4\rightarrow5)}$ represent phase delays imposed by the first and second metasurface layers, respectively (see SI Appendix Note 3). The insets reveal phase reconfiguration of the metasurface unit cell via resonator arm rotation, effectively switching the trainable meta-neurons from (M) to (I). (B) Confusion matrix generated from the testing data depicting the system's accuracy in the Iris problem. (C) Intensities (blue bars) and probabilities (red dashed lines) at the three detection units within the output layer for input samples corresponding to, from left to right, Setosa, Versicolor, and Virginica.}
\label{Fig6}
\end{figure*}
%%%%%%
It is critically important to note that metasurface-based neuromorphic systems developed to date, even those deployed in non-mechanical domains have not yet overcome the need to reconstruct their core wave-scattering components in order to retrain and carry out a different task. Whether it is TiO$_2$ pillars \cite{wu2020neuromorphic} or SiO$_2$-filled silicon slots \cite{fu2023photonic} which form the waveguide array, these geometries are fixed once manufactured. Therefore, owing to their inability to alter trainable nodes post design, the computational capacity of such systems remains limited to executing the single computational task that the constitutive cells were trained and physically constructed for. Motivated by the need to overcome this drawback, we demonstrate here the ability of the elastic neuromorphic metasurface to exercise an unprecedented degree of reconfigurability, allowing it to cater to distinct classification tasks, forgoing reconstruction and eliminating the need for costly remanufacturing. To illustrate, we re-utilize the baseline model generated for MNIST digit recognition and retrain it without changing any of its hyperparameters to execute the Iris flower classification task \cite{fisher1936use}. The new task involves the binning of Iris flower samples into three distinct classes (Setosa, Versicolor, and Virginica) based on four features: sepal length, sepal width, petal length, and petal width. The values of these features represent the amplitudes of four excitation point sources placed at the input layer (i.e., $K=4$), each two separated by $1.5$~m. The number of meta-neurons in the two metasurface layers is kept at $J=L=1,000$, while $N=3$ detection units corresponding to the three classes are defined at the output (readout) layer. The model is trained on $70\%$ of the available dataset ($105$ samples) and tested on the remaining $30\%$, yielding an accuracy of $97\%$ and $96\%$ for the training and testing datasets, respectively.

Figure~\ref{Fig6}A shows the precise phase profiles for all $1,000$ meta-neurons in the first (Spine $2 \to 3$) and second (Spine $4 \to 5$) metasurface layers for both the originally trained (MNIST, or M) and retrained (Iris, or I) datasets. The small insets reveal the phase reconfiguration of the metasurface unit cell via resonator arm rotation, effectively switching the trainable meta-neurons from (M) to (I). Figure~\ref{Fig6}B shows the confusion matrix for the 3 classes of the Iris dataset generated from the testing data. Furthermore, we show the intensity (blue bars) and probability (red dashed lines) distribution over the three detection units within the output (readout) layer in Fig.~\ref{Fig6}C, confirming the neuromorphic system's robust ability to relearn a diverse set of tasks within the same mechanical platform.
%%%%%%%%%%%%%%%%%%%%%%%%%%%%%%%%%%%%%%%%%%%%%%%%%%%%%%%%%%%%%%%%%%%%%%%%%%%%%%%%%%%%%%%%%%%%%%
\section{\label{sec:conc}CONCLUSIONS}
In this work, the foundations of metasurface-based neuromorphic computing have been introduced to the field of structural mechanics, utilizing guided in-plane vibrational waves in an elastic substrate. Edge excitations depicting the features of an input sample from two distinct multivariate datasets were fed to the neuromorphic system in the form of an array of spatially equidistant mechanical monopoles. These excitations were made to propagate through multiple layers of trained waveguides, eventually focusing the bulk portion of scattered wave intensity on the correct label identifier on a prescribed detection plane. Exploiting analogies between the different components of the proposed neuromorphic assembly and neural architectures, a customized neural network was defined accounting for non-trainable, constant wave propagation matrices (weights) and trainable phase delays (activation functions). \Revision{While the proposed neuromorphic system exhibits a high level of mechanical intelligence, it's equally important to consider its advantages against existing physical neural architectures. Diffractive neural networks, for example, share the principle of wave propagation through layered structures using passive optical components mimicking artificial neurons to perform intelligent tasks. However, their bulky designs and limited reconfigurability for different tasks pose challenges. MEMS-based neural networks, on the other hand, offer appealing features like adaptability to complex problems and low power consumption by mimicking biological neural networks. However, their reliance on microscopic moving parts introduces significant fabrication complexity and hinders miniaturization efforts.} Owing to the inherent reconfigurable nature of our physical meta-neurons within the metasurface arrays, the proposed system represents an embodiment of metasurface-based neuromorphic computers which are no longer constrained to a single unaltered task, and one where neurons can mechanically recompose their local phase profile to adapt to a new training scheme and execute a new task, on demand while maintaining a high performance accuracy. 

%%%%%%%%%%%%%%%%%%%%%%%%%%%%%%%%%%%%%%%%%%%%%%%%%%%%%%%%%%%%%%%%%%%%%%%%%%%%%%%%%%%%%%%%%%%%%%

\section*{Supplementary Material} See the supplementary material for a comprehensive understanding of the theoretical background of this work, and additional results referred
to in this manuscript.

%%%%%%%%%%%%%%%%%%%%%%%%%%%%%%%%%%%%%%%%%%%%%%%%%%%%%%%%%%%%%%%%%%%%%%%%%%%%%%%%%%%%%%%%%%%%%%

\begin{acknowledgments}
This work was supported by the US Army Research Office (ARO) and the Mechanical Behavior of Materials program under Grant W911NF-23-1-0078.
\end{acknowledgments}
%%%%%%%%%%%%%%%%%%%%%%%%%%%%%%%%%%%%%%%%%%%%%%%%%%%%%%%%%%%%%%%%%%%%%%%%%%%%%%%%%%%%%%%%%%%%%%

\section*{Data Availability} The data that supports the findings of this study are available within the article and its supplementary material.
%%%%%%%%%%%%%%%%%%%%%%%%%%%%%%%%%%%%%%%%%%%%%%%%%%%%%%%%%%%%%%%%%%%%%%%%%%%%%%%%%%%%%%%%%%%%%%

\section*{References}
\bibliography{references}% Produces the bibliography via BibTeX.

\end{document}